\documentclass{article}
\usepackage{threeparttable}

\usepackage{arxiv}

\usepackage[utf8]{inputenc} 
\usepackage[T1]{fontenc}    
\usepackage{nicefrac}       
\usepackage{microtype}      
\usepackage{lipsum}

\usepackage{graphicx}

\usepackage{latexsym}
\usepackage{textcomp}
\usepackage{longtable}
\usepackage{tabulary}
\usepackage{booktabs,array,multirow}
\usepackage{amsfonts,amsmath,amssymb}
\usepackage{natbib}
\usepackage{url}
\usepackage{hyperref}
\usepackage{etoolbox}
\usepackage{mathtools}
\usepackage{float}
\usepackage[toc,title,page]{appendix}
\usepackage{setspace}
\usepackage{subcaption}

\newif\iflatexml\latexmlfalse

\AtBeginDocument{\DeclareGraphicsExtensions{.pdf,.PDF,.eps,.EPS,.png,.PNG,.tif,.TIF,.jpg,.JPG,.jpeg,.JPEG}}

\usepackage[utf8]{inputenc}
\usepackage[english]{babel}

\usepackage{siunitx}

\iflatexml

\else

\author{
 Shuyi Liang \\
  University at Buffalo\\
   \And
 Takeshi Emura \\
  Institute of Statistical Mathematics
  \And
 Chang-Xing Ma \\
  University at Buffalo\\
  \And
  Yijing Xin \\
  The First Affiliated Hospital of Xiamen University\\
  \And
  Xin-Wei Huang\\
  University at Buffalo\\
  \texttt{xinweihuangstat@gmail.com}
}

\title{Testing the Homogeneity of Proportions for Correlated Bilateral Data via the Clayton Copula}

\begin{document}

\maketitle
\selectlanguage{english}

\begin{abstract}
Handling highly dependent data is crucial in clinical trials, particularly in fields related to  ophthalmology. Incorrectly specifying the dependency structure can lead to biased inferences. Traditionally, models rely on three fixed dependence structures, which lack flexibility and interpretation. In this article, we propose a framework using a more general model --- copulas --- to better account for dependency. We assess the performance of three different test statistics within the Clayton copula setting to demonstrate the framework's feasibility. Simulation results indicate that this method controls type I error rates and achieves reasonable power, providing a solid benchmark for future research and broader applications. Additionally, we present analyses of two real-world datasets as case studies.
\end{abstract}

\section{Introduction}

In randomized clinical trials (RCTs), researchers are typically interested in evaluating treatment effects across various groups. By comparing each treatment group to a control group, the treatment effect for each group can be assessed. Additionally, comparing differences among treatment groups allows for further inferences about new study or treatment designs. Therefore, understanding the differences between groups is especially important in RCTs. However, in ophthalmologic studies, samples are often not based solely on individual patients but rather on paired eye samples. Some characteristics may show in one or both eyes, which increases the complexity of the model. In traditional likelihood-based methods, different outcome combinations result in different likelihood contributions while similar considerations arise in studies involving other paired or grouped organs, such as ears, legs, or fingers. In the context that follows, we use ophthalmologic studies as an example to model and analyze bilateral eye data.

In bilateral RCT data, a frequently addressed issue is the dependency. Failure to account for the dependence will lead to biased statistical inference \citep{rosner1982statistical, dallal1988paired, donner1988analysis}. Naturally, we assume that the data from both eyes of the same sample exhibit a certain degree of dependence. For example, there are inherent differences in corneal thickness between individuals, which leads to noticeable distinctions in the design of myopic laser surgeries. However, the corneal thickness of both eyes within the same individual is usually very similar, so the surgical design for both eyes of an individual is often alike. Ignoring the similarity between both eyes when making inferences from RCT data can easily lead to an overestimation of statistical power. Therefore, appropriately considering dependence is very important. 

In previous studies, \citet{rosner1982statistical} proposed a "constant R" model to model intraclass correlation. The properties of this model and various related hypothesis testing methods were further investigated. For example, \citet{ma2015homogeneity} developed statistical tests for the homogeneity of multiple treatment groups. When the sample size is limited, exact methods were proposed by \citet{liu2017exactR}. However, the "constant R" model performed poorly because the characteristic is almost certain to occur bilaterally with widely varying group-specific prevalence. \citet{dallal1988paired} developed a model that further employed a more flexible conditional probability assumption with group differences. \citet{li2020statistical} proposed asymptotic and exact methods for testing the homogeneity of the conditional probabilities among multiple groups.  On the other hand, \citet{donner1989statistical} made assumptions based on a fixed value of correlation and proposed an adjusted Pearson's chi-square test to test the differences among proportions arising from eye-specific binary data. Base on Donner's model, homogeneity tests for comparing multiple treatment groups using asymptotic and exact methods were investigated by \citet{ma2017testing} and \citet{liu2017exactrho}, respectively. 

In recent years, both clinical trials and observational studies have produced a growing amount of paired data. Many researchers have realized that paired data differ from other data types. As a result, this field has gained considerable attention and importance. Over a relatively short period, numerous studies examining different assumptions have emerged based on the three fundamental approaches. \citet{liang2024homogeneity} studied homogeneity tests of the risk difference between two groups across different strata and proposed confidence interval construction of the risk difference when there is no stratification effect, and \citet{tian2024testing} proposed three statistical tests for assessing stratification effects on the risk ratio of two groups under Dallal's model. \citet{zhang2023testing} studied stratification effect on the risk difference between two groups under Donner's model. Homogeneity tests of odds ratio was also examined by \citet{hua2024testing} under different strata. \citet{zhang2024simultaneous} investigated many-to-one simultaneous confidence intervals of risk ratios.

We recognize that the correlation assumptions in these models are based on linear relationships. However, binary outcomes in eye data may involve more complex and potentially nonlinear dependence structures concerning proportion rates. Based on this, a more general framework capable of handling both linear and nonlinear dependencies warrants consideration. Copula functions are one of the popular tools of addressing both linear and nonlinear dependencies. Copula is also powerful to offer strong interpretative ability. It is widely employed in various research areas within clinical trials and observational studies, for instance, \citet{emura2016gene}, \citet{emura2017joint}, and \citet{huang2021copula} applied bivariate copula functions to survival analysis for modeling semi-competing risks data. Modeling discrete data has been explored by \citet{panagiotelis2012pair}, \citet{koopman2018dynamic}, and \citet{huang2024computational}. Binary outcome models, a special case of discrete margin models, have been investigated in regression applications \citep{radice2016copula, mesfioui2023copula}, but their use in clinical data remains limited. Some well-established examples of the corresponding tools such as goodness-of-fit tests \citep{wang2003estimating, huang2021model} and tests for symmetry \citep{jaser2021tests, lyu2023testing} are available for reference. The prevalence and versatility of copula models allow this problem to be simplified to how to utilize copulas for handling binary outcome RCT data.

In this context, a copula can be understood as the underlying dependence structure of the treatment success rate/incidence rate model for both eyes. For example, it can be a linear correlation, such as the traditional Pearson's $\rho$ model. Alternatively, copulas can allow tail dependence where there is strong dependency at low treatment success rates, but the dependency decreases as the success rate increases. Copulas, as a collection of these potential patterns, offer numerous options. Furthermore, the outcome data, that is, the marginal binary data, appear as paired samples representing these success rate/incidence rate models, containing potential dependence  information. Therefore, it is reasonable to use copula models to generalize this application.

Our contributions are as follows:
\begin{enumerate}
    \item We propose a novel framework to handle dependent binary outcomes of eyes in ophthalmology trials. This framework integrates a wide range of linear and nonlinear dependence structures by incorporating copulas, thereby achieving excellent flexibility. Several classical approaches \citep{rosner1982statistical, dallal1988paired, donner1989statistical} can be generalized by our framework.
    \item The proposed framework standardizes the correlation measure to Kendall's tau under Archimedean copulas, facilitating the comparison of correlations across models.
    \item We provide an example of performing likelihood inference and hypothesis testing for the treatment success rate parameters under our copula model. This inference forms the fundamental basis for future research on model comparison and goodness-of-fit test.
\end{enumerate}

We primarily discuss the copula model and the rationale for using copulas with binary outcomes in Section \ref{sec:copula model}. Sections \ref{sec:likelihood} and \ref{sec:testing} provide technical details on likelihood inference and test statistics. In Section \ref{sec:simulation}, we outline the simulation design used to evaluate the statistical power of various test statistics within our proposed model. Section \ref{sec:data} presents a data analysis illustrating the model. Finally, in Section \ref{sec:discussion}, we discuss our findings and outline future work based on this framework.

\section{Method} \label{sec:proposed_method}

We consider an ophthalmology dataset comparing $g$ groups of individuals, each with paired eye information. Group $i$ consists of $m_i$ individuals, totaling $N = \sum_{i=1}^{g} m_i$ observations. For each individual $j$ in group $i$, the status of the $k$th eye is represented by $Z_{ijk}$, where $Z_{ijk} = 1$ indicates a diseased eye and $Z_{ijk} = 0$ denotes a healthy eye, with $k = 1, 2$. To summarize the data (see Table \ref{tb:tb1}), let $m_{i\ell} = \sum_{j=1}^{m_i} \mathbb{I}(Z_{ij1} + Z_{ij2} = \ell)$ represent the number of individuals with exactly $\ell$ diseased eyes in group $i$, and let $S_\ell = \sum_{i=1}^{g} m_{i\ell}$ denote the total number of individuals with $\ell$ diseased eyes across all groups.

\begin{table}[h]
\centering
\caption{Frequencies of the Number of Diseased Eyes}
\label{tb:tb1}
\begin{tabular}{lccccc}
\hline
Number of Diseased Eyes & Group 1 & Group 2 & \dots & Group $g$ & Total \\
\hline
0 & $m_{10}$ & $m_{20}$ & \dots & $m_{g0}$ & $S_0$ \\
1 & $m_{11}$ & $m_{21}$ & \dots & $m_{g1}$ & $S_1$ \\
2 & $m_{12}$ & $m_{22}$ & \dots & $m_{g2}$ & $S_2$ \\
\hline
\textbf{Total} & $m_1$ & $m_2$ & \dots & $m_g$ & $N$ \\
\hline
\end{tabular}
\end{table}

\subsection{Copula Model} \label{sec:copula model}

The term “copula” originates from Latin, meaning link, tie, or bond. A bivariate copula is a function that links two uniform marginal distributions such that
$$
\Pr(U \le u, V \le v) = C_\theta(u, v),
$$
where $U$ and $V$ are the uniform random variables on $[0,1]$, and $\theta$ is the copula parameter. The joint distribution function satisfies properties as the usual joint probability such as $C(u, 1) = u$, $C(1, v) = v$, and $C(u, 0) = C(0, v) = 0$. Copulas are advantageous thanks to Sklar’s theorem \cite{sklar1959fonctions}, which allows for different marginal distributions, enabling for arbitrary random variables $X$ and $Y$ such that
$$
\Pr(X \le x, Y \le y) = C_\theta(F_X(x), F_Y(y)).
$$
Notably, $C_\theta$ is unique when the marginals are continuous. Further details can be found in \citet{nelsen2006introduction}.

Let $\Pr(Z_{ijk} = 1) = \pi_{ik}$ denote the underlying treatment success rate of the $k$th eye in group $i$. The paired eye data can be modeled as
$$
\Pr(Z_{ij1} \le z_{ij1}, Z_{ij2} \le z_{ij2}) = C_\theta\bigl(F_{i1}(z_{ij1}), F_{i2}(z_{ij2})\bigr),
$$
where $C_\theta$ is a copula function. The marginal distributions follow a Bernoulli CDF:
\begin{gather*}
    F_{ik}(z_{ijk}) =
\begin{cases}
1 - \pi_{ik}, & \text{if } 0\leq z_{ijk} < 1, \\
1, & \text{if } z_{ijk} \geq 1.
\end{cases}
\end{gather*}

Using copula properties, we can simplify the model algebraically. In total, there are four joint probabilities:
\begin{enumerate}
    \item $\Pr(Z_{ij1} = 0, Z_{ij2} = 0) = C_\theta\bigl(1 - \pi_{i1}, 1 - \pi_{i2}\bigr) \overset{*}{=} C_\theta(1 - \pi_i, 1 - \pi_i)$,
    \item $\Pr(Z_{ij1} = 1, Z_{ij2} = 0) = 1 - \pi_{i2} - C_\theta\bigl(1 - \pi_{i1}, 1 - \pi_{i2}\bigr) \overset{*}{=} 1 - \pi_i - C_\theta(1 - \pi_i, 1 - \pi_i)$,
    \item $\Pr(Z_{ij1} = 0, Z_{ij2} = 1) = 1 - \pi_{i1} - C_\theta\bigl(1 - \pi_{i1}, 1 - \pi_{i2}\bigr) \overset{*}{=} 1 - \pi_i - C_\theta(1 - \pi_i, 1 - \pi_i)$,
    \item $\Pr(Z_{ij1} = 1, Z_{ij2} = 1) = 1 - \bigl(1 - \pi_{i1}\bigr) - \bigl(1 - \pi_{i2}\bigr) + C_\theta\bigl(1 - \pi_{i1}, 1 - \pi_{i2}\bigr) \overset{*}{=} 1 - 2\bigl(1 - \pi_i\bigr) + C_\theta(1 - \pi_i, 1 - \pi_i)$.
\end{enumerate}
Without loss of generality, we use the operation $\overset{*}{=}$ to emphasize the simplification for $\pi_{i1} = \pi_{i2} = \pi_i$. This assumption will be hold in the following discussion.
    
\begin{figure}[h]
\centering
\includegraphics[width=\textwidth]{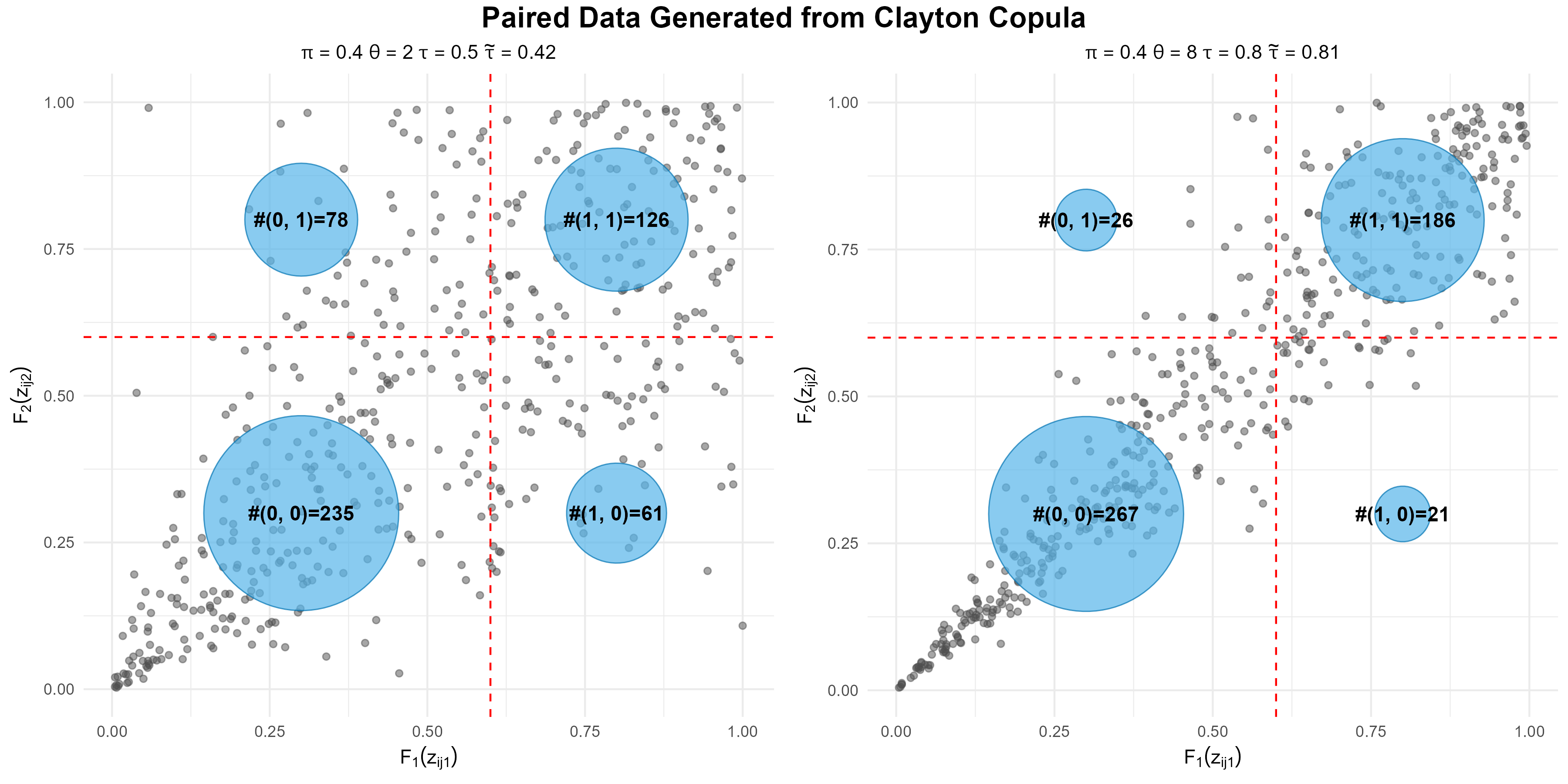} 
\caption{Example of the Clayton Copula. The data points are generated from the Clayton copula. The dashed lines ($F_1 = 0.6$ and $F_2 = 0.6$) indicates the success treatment rate is $\pi_i = 0.4$. Any value over these lines is categorized as $z_{ijk}=1$. Points are classified to the four different cases. The frequencies four each cases are shown by the bubbles.}
\label{fg:clayton}
\end{figure}

The main advantage of using copulas is their flexibility in modeling dependence structures. Various copula families can be applied based on data patterns. For example, the Clayton copula is appropriate for lower-tail dependence, while the Joe copula is well-suited for upper-tail dependence. In Figure \ref{fg:clayton}, we generate paired Bernoulli data using the Clayton copula,
\begin{equation}
    \label{eq:clayton}
    C_\theta(u, v) = \bigl(u^{-\theta} + v^{-\theta} - 1\bigr)^{-1/\theta}, \theta>0,
\end{equation}

for strong ($\tau = 0.8$) and moderate ($\tau = 0.5$) dependency with a marginal treatment success rate of 0.4. The points in the figures represent marginal uniform variables that reflect the copula’s correlation structure. Since the marginals are binary outcomes, a bubble plot effectively illustrates the frequencies of each case, where bubble sizes indicate the counts. The Clayton copula pattern shows stronger correlations for both successes treated eyes compared to both failure treated eyes. Cases with one success and one failure treated eye are less frequent under stronger correlation, and all data points cluster more closely around the center line.

The second advantage is the straightforward interpretation of the dependent parameters. Models proposed by \citet{donner1989statistical}, \citet{rosner1982statistical}, and \citet{dallal1988paired} each provide different interpretations of dependency. Copula models offers Kendall's tau measure that is free from the marginal distributions. Specifically, for Archimedean copulas, Kendall's $ \tau $ can be easily obtained from the generator functions \cite{genest1993statistical}. In the Clayton copula, $ \tau = \frac{\theta}{\theta + 2} $. Kendall’s $ \tau $ is a unified measure of concordant pairs, thus the dependence is comparable between models. With the help of visualizing copulas, it offers a deeper understanding of dependence by capturing the specific structure beyond a single numerical value. The Pearson correlation coefficient between the treatment rates of two eyes is also available such that $\rho_i=\frac{C_\theta(1 - \pi_i, 1 - \pi_i)-(1-\pi_i)^2}{\pi_i(1-\pi_i)}$, though it depends on the marginal distributions.

Our proposed copula model also generalizes several classical models. For example, \citet{rosner1982statistical} assumed
$
\Pr(Z_{ij1} = 1 \mid Z_{ij2} = 1) = R\pi_i.
$
where $R\pi_i$ is the treatment success rate. We can show that this is equivalent to the model with
$$
C_\theta(1 - \pi_i, 1 - \pi_i) = R\pi_i^2 - 2\pi_i + 1.
$$
This property holds in the independence case when \( R = 1 \), where $
C_\theta(1 - \pi_i, 1 - \pi_i) = (1 - \pi_i)^2,
$ which implies an independence copula. \citet{donner1988analysis}'s model considered a common correlation coefficient $\rho = \text{Corr}(Z_{ij1}, Z_{ij2})$. In the same concept, this model is denoted by copula as $$C_\theta(1 - \pi_i, 1 - \pi_i) = (1-\rho)\pi_i^2 + (\rho-2)\pi_i + 1.$$ The same generalization is applied to \citet{dallal1988paired}'s model $\Pr(Z_{ij1} = 1 \mid Z_{ij2} = 1) = \gamma_i$ leads to a copula $$C_\theta(1 - \pi_i, 1 - \pi_i) = (\gamma_i-2)\pi_i+1.$$ These results also reveal the relationship between these three classical models.

\subsection{Likelihood Inference} \label{sec:likelihood}
We are interested in testing $H_0:$ $\pi_1=\pi_2=...\pi_g=\pi_0$ versus $H_a:$ $\pi_i \neq \pi_j$ for some $i \neq j$, where $\pi_0$ is an unknown quantity. Let \( \tilde{M} = (m_{10}, \dots, m_{g0}, m_{11}, \dots, m_{g1}, m_{12}, \dots, m_{g2}) \) represent the observed summarized data. Assuming equal disease rates \( \pi_i \) within group \( i \), the log-likelihood function is expressed as:
\begin{align*}
    \ell(\pi_1, \dots, \pi_g, \theta \mid \tilde{M}) &= \sum_{i=1}^{g} \Bigg[ 
        m_{i0} \log\left(C_\theta(1 - \pi_i, 1 - \pi_i)\right) \\
        &\quad +\, m_{i1} \log\left(2(1 - \pi_i) - 2C_\theta(1 - \pi_i, 1 - \pi_i)\right) \\
        &\quad +\, m_{i2} \log\left(1 - 2(1 - \pi_i) + C_\theta(1 - \pi_i, 1 - \pi_i)\right) .
    \Bigg]
\end{align*}

We apply the Clayton copula for inference. The Clayton copula function is defined as Equation \ref{eq:clayton} with $\theta >0$, thus the log-likelihood under the alternative becomes:
\begin{align*}
    \ell(\pi_1, \dots, \pi_g, \theta \mid \tilde{M}) &= \sum_{i=1}^{g} \Bigg[
        m_{i0} \log\left(\left(2(1 - \pi_i)^{-\theta} - 1\right)^{-\frac{1}{\theta}}\right) \\
        &\quad +\, m_{i1} \log\left(2(1 - \pi_i) - 2\left(2(1 - \pi_i)^{-\theta} - 1\right)^{-\frac{1}{\theta}}\right) \\
        &\quad +\, m_{i2} \log\left(1 - 2(1 - \pi_i) + \left(2(1 - \pi_i)^{-\theta} - 1\right)^{-\frac{1}{\theta}}\right).
    \Bigg]
\end{align*}
Note that $1-2\pi_i \leq C_{\theta} \leq 1-\pi_i$ since $\Pr(Z_{ij1} = 0, Z_{ij2} = 0)$, $\Pr(Z_{ij1} = 0, Z_{ij2} = 1)$, $\Pr(Z_{ij1} = 1, Z_{ij2} = 0)$, and $\Pr(Z_{ij1} = 1, Z_{ij2} = 1)$ fall within the interval [0, 1]. It is straightforward to verify that the Clayton copula is an increasing function of $\theta$ given fixed $\pi_i$. Therefore, the following inequality holds:
\[ \lim_{\theta \to0} (2(1-\pi_i)^{-\theta}-1)^{-\frac{1}{\theta}} \leq C_{\theta} \leq \lim_{\theta \to\infty} (2(1-\pi_i)^{-\theta}-1)^{-\frac{1}{\theta}}, \]
which is equivalent to
\[ (1-\pi_i)^2 \leq C_{\theta} \leq 1-\pi_i . \]
It is clear that any $\theta>0$ satisfies the condition that $1-2\pi_i \leq C_{\theta} \leq 1-\pi_i$.

Under $H_0$, the log likelihood function  can be expressed as:
\begin{align*}
\ell_0(\pi_0, \theta \mid \tilde{M}) &= \sum_{i=1}^{g} \Bigg[
  m_{i0} \log\left(\left(2(1 - \pi_0)^{-\theta} - 1\right)^{-\frac{1}{\theta}}\right) \\
  &\quad +\, m_{i1} \log\left(2(1 - \pi_0) - 2\left(2(1 - \pi_0)^{-\theta} - 1\right)^{-\frac{1}{\theta}}\right) \\
  &\quad +\, m_{i2} \log\left(1 - 2(1 - \pi_0) + \left(2(1 - \pi_0)^{-\theta} - 1\right)^{-\frac{1}{\theta}}\right)
  \Bigg].
\end{align*}
To obtain maximum likelihood estimators (MLEs) of $\pi_i$ and $\theta$, the Newton-type algorithm provided in the R function \verb|nlm()| can be used \citep{dennis1996numerical, schnabel1985modular} since there is no analytical form of solutions. The algorithm finds a local minimizer of a real valued function without any constrains. However, all parameters in the proposed model are not necessarily from the real line. Hence, two transformations, $\log \left( \frac{\pi_i}{1-\pi_i} \right)$ and $\log(\theta)$, are employed to avoid introducing additional constrains. The final minimizer can be restored via a simple inverse of these functions. Denote the MLEs of $\pi_i$ and $\theta$ as $\hat{\pi}_i$ and $\hat{\theta}$ under $H_a$, respectively. Under the null hypothesis, the MLEs of $\pi_0$ and $\theta$ are represented by $\hat{\pi}_{H_0}$ and $\hat{\theta}_{H_0}$, respectively.

\subsection{Hypothesis testing} \label{sec:testing}
\subsubsection{Likelihood Ratio Test ($T_{LR}$)}
The likelihood ratio (LR) test statistic, denoted as $T_{LR}$, is given by:
\[T_{LR}=2(\ell(\hat{\pi}_1, \dots, \hat{\pi}_g, \hat{\theta} \mid \tilde{M})-\ell_0(\pi_0, \hat{\theta} \mid \tilde{M})).\]
Under $H_0$, $T_{LR}$ follows asymptotically a chi-square distribution with $g-1$ degrees of freedom \citep{wilks1938large}.

\subsubsection{Score Test ($T_{S}$)}
Let $U=(\frac{\partial \ell}{\partial \pi_1}, \frac{\partial \ell}{\partial \pi_2}, ...,\; \frac{\partial \ell}{\partial \pi_g}, \frac{\partial \ell}{\partial \theta} )^T$ and $I$ be the Fisher information matrix. The score test statistic is defined by:
\[T_S=U^TI^{-1}U\big| \pi_1=\pi_2=...\pi_g=\hat{\pi}_{H_0},\theta=\hat{\theta}_{H_0}, \]
where 
\[
I=
\begin{bmatrix}
I_{11} & 0 & & &0 & I_{1(g+1)}\\
0 & I_{22} &  & & 0& I_{2(g+1)}\\
   & & & \ddots & &\\
 0  &0 & & &I_{gg} &I_{g(g+1)}\\
I_{(g+1)1}  &  I_{(g+1)2} & & &I_{(g+1)g} &I_{(g+1)(g+1)}
\end{bmatrix}_{(g+1)\times (g+1).}
\]
The expression of $I_{ii}$, $I_{i(g+1)}$, and $I_{(g+1)(g+1)}$ ($i=1,2,...,g$) can be found in the Appendix. The score test statistic $T_S$ can be further simplified as:
\[I=\frac{-d_i(-\sum_{j \neq i}^{g} \frac{d_i I^2_{(g+1)j}}{I_{ii}I_{jj}} +\sum_{j \neq i}^{g}\frac{I_{(g+1)i}d_j I_{(g+1)j}}{I_{ii}I_{jj}}  + \frac{d_iI_{(g+1)(g+1)}}{I_{ii}} )}{\sum_{i=1}^{g}\frac{I^2_{(g+1)1}}{I_{ii}}-I_{(g+1)(g+1)}},\]
where 
\begin{flalign*}
d_i 
=\;
-\;m_{\textrm{i1}}\,
\frac{
\displaystyle \left(\, 4/\omega(\pi_i)^{\tfrac{1}{\theta}+1}\,(1-\pi_i)^{\theta+1}\;-\;2\right)
}{
\displaystyle 2\,\pi_i \;+\;{2}/{\omega(\pi_i)^{\tfrac{1}{\theta}}}\;-\;2
}
\;-\;
m_{\textrm{i2}}\,
\frac{
\displaystyle \left(\,{2}/{\omega(\pi_i)^{\tfrac{1}{\theta}+1}\,(1-\pi_i)^{\theta+1}}\;-\;2\right)
}{
\displaystyle 2\,\pi_i \;+\;{1}/{\omega(\pi_i)^{\tfrac{1}{\theta}}}\;-\;1
}
\;-\;
\frac{
2\,m_{\textrm{i0}}
}{
2 - 2\,\pi_i \;-\; \bigl(1-\pi_i\bigr)^{\theta+1}
}\, ,\\
\text{with } \omega(\pi_i) \;=\; {2}/{\bigl(1-\pi_i\bigr)^\theta} \;-\; 1.
\end{flalign*}

$T_S$ is asymptotically distributed as a chi-square distribution with $g-1$ degrees of freedom according to \citet{rao1948large}.

\subsubsection{Wald Test ($T_{W}$)}
To construct the Wald-type test, we first define $\beta=(\pi_1,\pi_2,...,\pi_g,\theta)^T$ and note that the null hypothesis $H_0$ is equivalent to
\[C \beta=
\begin{bmatrix}
1 & -1 & & & &0\\
 & 1 & -1 & & &0\\
 & & \ddots & \ddots & &\vdots\\
 & & & 1 & -1 & 0
\end{bmatrix}_{(g-1)\times(g+1)}\times
\begin{bmatrix}
\pi_1\\
\pi_2\\
\vdots\\
\pi_g\\
\theta
\end{bmatrix}=
\begin{bmatrix}
0\\
0\\
\vdots\\
0
\end{bmatrix}_{(g-1)\times 1}.
\]
Then the Wald test statistic $T_W$ can be written as:
\[T_W=(C\beta)^T(CI^{-1}C^T)^{-1}(C\beta)|\pi_1=\hat{\pi}_1,\pi_2=\hat{\pi}_2,...,\pi_g=\hat{\pi}_g,\theta=\hat{\theta},\]
which follows asymptotically a $\mathcal{X}^2$ distribution with $g-1$ degrees of freedom under the alternative hypothesis $H_a$ \citep{wald1943tests}.

\section{Simulation} \label{sec:simulation}
\begin{figure}[H]
     \centering
\includegraphics[width=\textwidth]{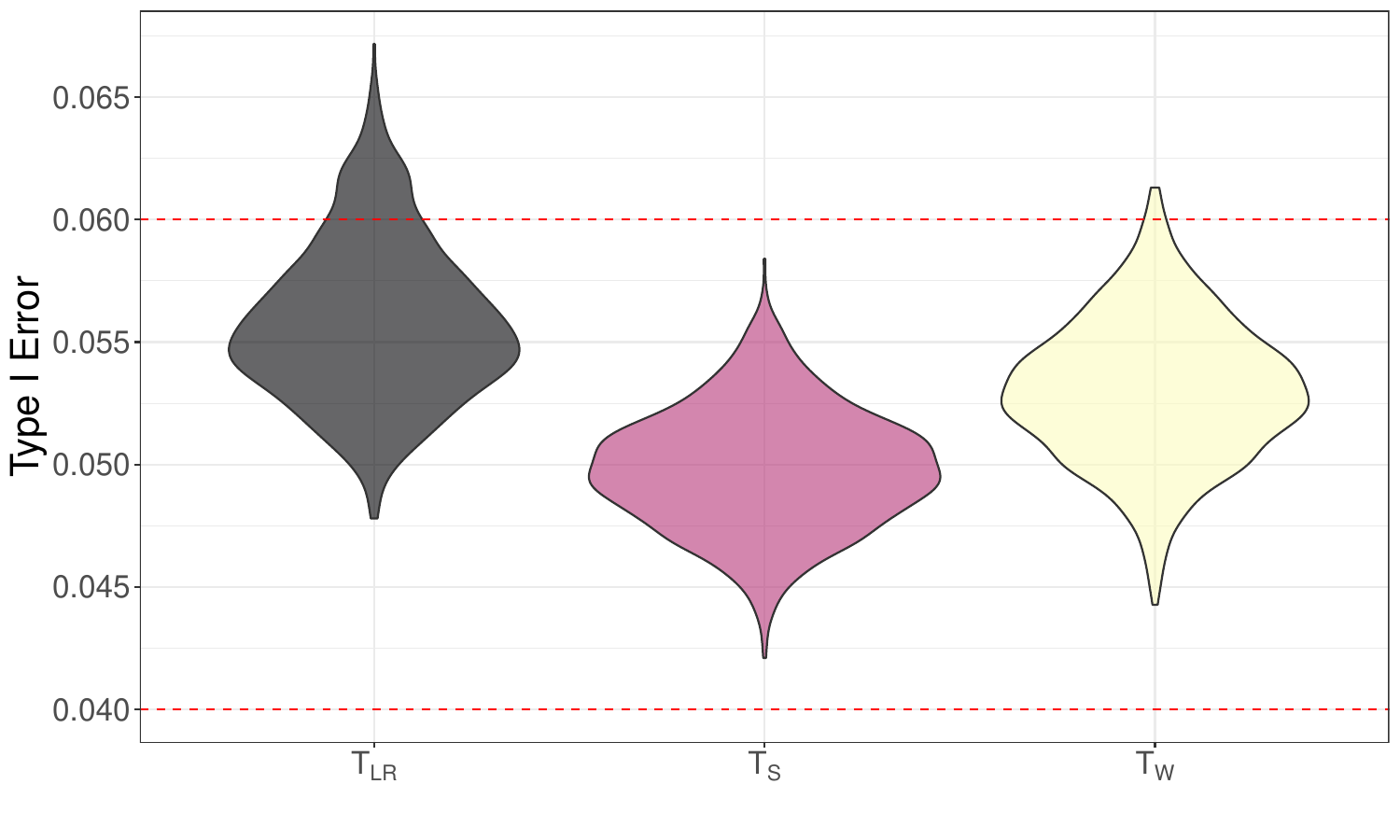}
\caption{Violin Plots of Empirical Type I Errors for $g=3$ and $m_i=30$}
        \label{fig:violin_g3_m_30}
\end{figure}

\begin{figure}[H]
     \centering
\includegraphics[width=\textwidth]{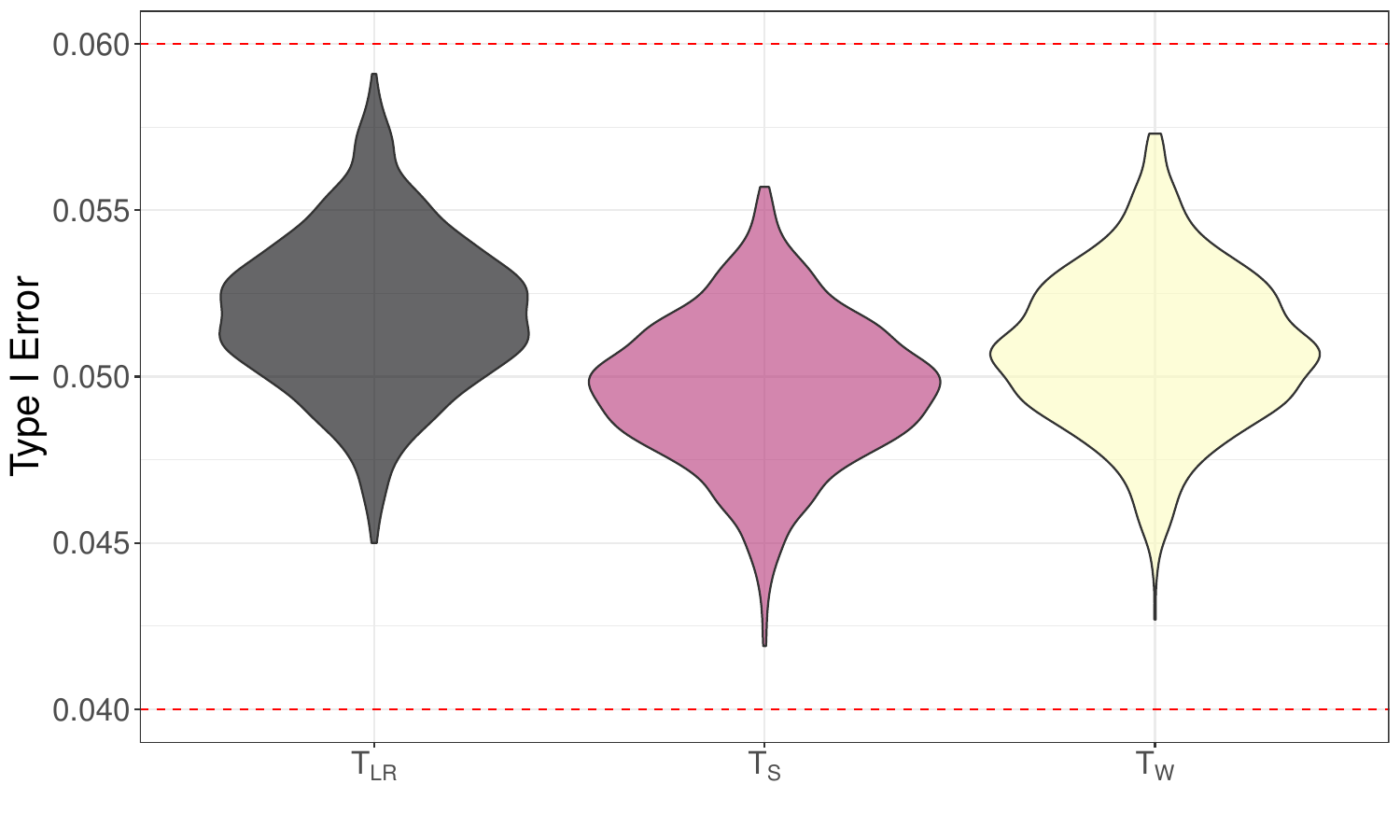}
        \caption{Violin Plots of Empirical Type I Errors for $g=6$ and $m_i=100$}
        \label{fig:violin_g6_m_100}
\end{figure}

The type I errors (TIEs) and statistical powers of the three tests (LR test, score test, and Wald test) under our proposed copula model are investigated in this section using various parameter settings. Given a balanced design with the sample size $m_i=30$, $55$, and $100$, a random sample is drawn from different combination of $\pi_i$ and $\theta$, where $\pi_i \in \{0.4,0.5,0.6,0.7\}$ and $\theta \in \{0,2,8\}$. The higher the value of $\theta$, the stronger the correlation between the two eyes. Note that $\theta=0$ gives independence between two eyes. The empirical type I errors are determined by a total of 10,000 simulations and are presented in \hyperref[tab:TIE3]{Table \ref*{tab:TIE3}} and \hyperref[tab:TIE6]{Table \ref*{tab:TIE6}} under the group number $g=3$ and 6, respectively. To cover a broader parameter space, we first randomly generate a combination of $\pi_i$ and $\theta$ that makes $\Pr(Z_{ij1} = 0, Z_{ij2} = 0),\; \Pr(Z_{ij1} = 0, Z_{ij2} = 0),\; \Pr(Z_{ij1} = 0, Z_{ij2} = 0)$, and $\Pr(Z_{ij1} = 0, Z_{ij2} = 0)$ fall within the interval (0.1, 1]. The rationale of adding this restriction is to avoid sparse tables that lead to irregular behaviors of MLEs and test statistics. After that, ten thousand simulations are conducted to evaluate the empirical TIE. The aforementioned process is repeated 1,000 times and the results are presented in \hyperref[fig:violin_g3_m_30]{Figure \ref*{fig:violin_g3_m_30}} to \hyperref[fig:violin_g6_m_100]{Figure \ref*{fig:violin_g6_m_100}}.

\begin{singlespace}
\begin{center}
\addtolength{\tabcolsep}{0.1em}
\begin{longtable}[H]{c c c | c c c | c c c | c c c }
\caption{The Empirical Type I Errors (\%) when $g=3$ under the Nominal Level $\alpha=5\%$ } \label{tab:TIE3} \\

\hline
 &  &  & \multicolumn{3}{c|}{$m_i=30$} & \multicolumn{3}{c|}{$m_{i}=55$} & \multicolumn{3}{c}{$m_{i}=100$}  \\ \cline{4-12}
$\theta$ & $\pi_i$ & $\rho_i$ & $T_{LR}$ & $T_{S}$ & $T_W$ & $T_{LR}$ & $T_{S}$ & $T_W$ & $T_{LR}$ & $T_{S}$ & $T_W$   \\  \hline

\endfirsthead

\multicolumn{12}{l}%
{{ \tablename\ \thetable{}: (Continued)}} \\
\hline
 &  &  & \multicolumn{3}{c|}{$m_i=30$} & \multicolumn{3}{c|}{$m_{i}=55$} & \multicolumn{3}{c}{$m_{i}=100$}  \\ \cline{4-12}
$\theta$ & $\pi_i$ & $\rho_i$ & $T_{LR}$ & $T_{S}$ & $T_W$ & $T_{LR}$ & $T_{S}$ & $T_W$ & $T_{LR}$ & $T_{S}$ & $T_W$   \\  \hline
\endhead

\hline \multicolumn{12}{c}{{Continued on next page}} \\ \hline
\endfoot

\hline \hline
\endlastfoot

0 & 0.4 & 0 & 4.760 & 4.110 & 4.215 & 5.220 & 4.670 & 4.677 & 5.400 & 4.980 & 5.047 \\ 
        ~ & 0.5 & 0 & 4.880 & 4.140 & 4.182 & 4.800 & 4.390 & 4.466 & 5.000 & 4.580 & 4.635 \\ 
        ~ & 0.6 & 0 & 5.030 & 4.320 & 4.295 & 5.030 & 4.690 & 4.783 & 5.130 & 4.750 & 4.787 \\ 
        ~ & 0.7 & 0 & 4.940 & 4.110 & 4.136 & 5.000 & 4.430 & 4.503 & 5.070 & 4.520 & 4.589 \\ \hline
2 & 0.4 & 0.452 & 5.980 & 5.440 & 5.840 & 4.920 & 4.670 & 4.900 & 5.630 & 5.560 & 5.590 \\ 
        ~ & 0.5 & 0.512 & 5.580 & 5.090 & 5.360 & 4.900 & 4.750 & 4.800 & 4.880 & 4.740 & 4.850 \\ 
        ~ & 0.6 & 0.562 & 5.420 & 5.080 & 5.220 & 5.260 & 5.080 & 5.150 & 4.680 & 4.620 & 4.700 \\ 
        ~ & 0.7 & 0.605 & 5.480 & 5.110 & 5.140 & 5.300 & 5.120 & 5.130 & 5.000 & 4.930 & 4.900 \\ \hline
8 & 0.4 & 0.795 & 5.771 & 5.351 & 5.561 & 5.150 & 4.890 & 4.980 & 4.900 & 4.850 & 4.880 \\ 
        ~ & 0.5 & 0.834 & 5.094 & 4.783 & 4.833 & 5.190 & 5.130 & 5.150 & 5.040 & 4.900 & 4.970 \\ 
        ~ & 0.6 & 0.862 & 5.367 & 5.097 & 5.067 & 5.370 & 5.310 & 5.300 & 5.020 & 4.920 & 4.940 \\ 
        ~ & 0.7 & 0.881 & 5.481 & 5.350 & 5.229 & 5.031 & 4.800 & 4.800 & 5.250 & 5.020 & 5.010 \\

\end{longtable}
\end{center}
\end{singlespace}

\begin{singlespace}
\begin{center}
\addtolength{\tabcolsep}{0.1em}
\begin{longtable}[H]{c c c | c c c | c c c | c c c }
\caption{The Empirical Type I Errors (\%) when $g=6$ under the Nominal Level $\alpha=5\%$ } \label{tab:TIE6} \\

\hline
 &  &  & \multicolumn{3}{c|}{$m_i=30$} & \multicolumn{3}{c|}{$m_{i}=55$} & \multicolumn{3}{c}{$m_{i}=100$}  \\ \cline{4-12}
$\theta$ & $\pi_i$ & $\rho_i$ & $T_{LR}$ & $T_{S}$ & $T_W$ & $T_{LR}$ & $T_{S}$ & $T_W$ & $T_{LR}$ & $T_{S}$ & $T_W$   \\  \hline

\endfirsthead

\multicolumn{12}{l}%
{{ \tablename\ \thetable{}: (Continued)}} \\
\hline
 &  &  & \multicolumn{3}{c|}{$m_i=30$} & \multicolumn{3}{c|}{$m_{i}=55$} & \multicolumn{3}{c}{$m_{i}=100$}  \\ \cline{4-12}
$\theta$ & $\pi_i$ & $\rho_i$ & $T_{LR}$ & $T_{S}$ & $T_W$ & $T_{LR}$ & $T_{S}$ & $T_W$ & $T_{LR}$ & $T_{S}$ & $T_W$   \\  \hline
\endhead

\hline \multicolumn{12}{c}{{Continued on next page}} \\ \hline
\endfoot

\hline \hline
\endlastfoot

0 & 0.4 & 0 & 4.910 & 4.310 & 4.396 & 4.910 & 4.540 & 4.651 & 5.060 & 4.870 & 4.854 \\ 
        ~ & 0.5 & 0 & 4.960 & 4.550 & 4.545 & 4.640 & 4.510 & 4.543 & 5.050 & 4.660 & 4.699 \\ 
        ~ & 0.6 & 0 & 5.100 & 4.380 & 4.445 & 4.490 & 4.420 & 4.430 & 4.820 & 4.540 & 4.611 \\ 
        ~ & 0.7 & 0 & 5.070 & 4.220 & 4.364 & 4.770 & 4.540 & 4.522 & 5.340 & 5.090 & 5.097 \\ 
2 & 0.4 & 0.452 & 5.360 & 4.720 & 5.110 & 5.300 & 4.990 & 5.040 & 5.130 & 4.820 & 5.020 \\ 
        ~ & 0.5 & 0.512 & 5.140 & 4.500 & 4.920 & 5.340 & 4.970 & 5.220 & 5.480 & 5.320 & 5.370 \\ 
        ~ & 0.6 & 0.562 & 5.870 & 5.300 & 5.620 & 5.400 & 5.080 & 5.220 & 5.450 & 5.240 & 5.350 \\ 
        ~ & 0.7 & 0.605 & 5.950 & 5.280 & 5.550 & 5.290 & 5.010 & 5.070 & 4.840 & 4.760 & 4.790 \\ 
8 & 0.4 & 0.795 & 5.570 & 4.840 & 5.140 & 5.320 & 4.980 & 5.030 & 4.970 & 4.790 & 4.920 \\ 
        ~ & 0.5 & 0.834 & 5.710 & 5.200 & 5.340 & 5.200 & 4.810 & 4.850 & 5.040 & 4.950 & 4.920 \\ 
        ~ & 0.6 & 0.862 & 5.160 & 4.650 & 4.700 & 5.400 & 5.120 & 5.120 & 4.910 & 4.760 & 4.730 \\ 
        ~ & 0.7 & 0.881 & 5.011 & 4.420 & 4.430 & 5.200 & 4.940 & 4.940 & 5.100 & 4.970 & 4.970 \\

\end{longtable}
\end{center}
\end{singlespace}

\begin{figure}[H]
     \centering
\includegraphics[width=\textwidth]{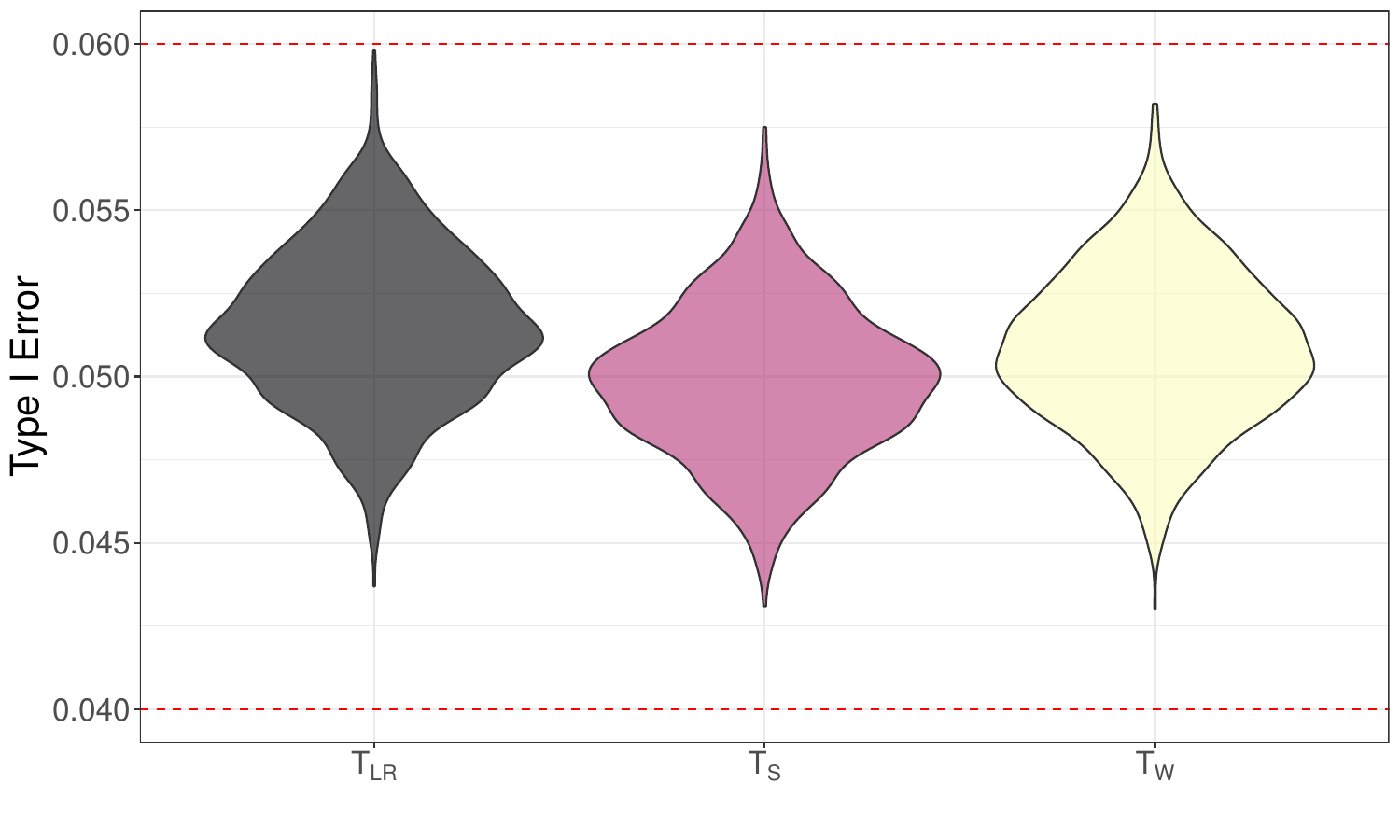}
        \caption{Violin Plots of Empirical Type I Errors for $g=3$ and $m_i=100$}
        \label{fig:violin_g3_m_100}
\end{figure}

\begin{figure}[H]
     \centering
\includegraphics[width=\textwidth]{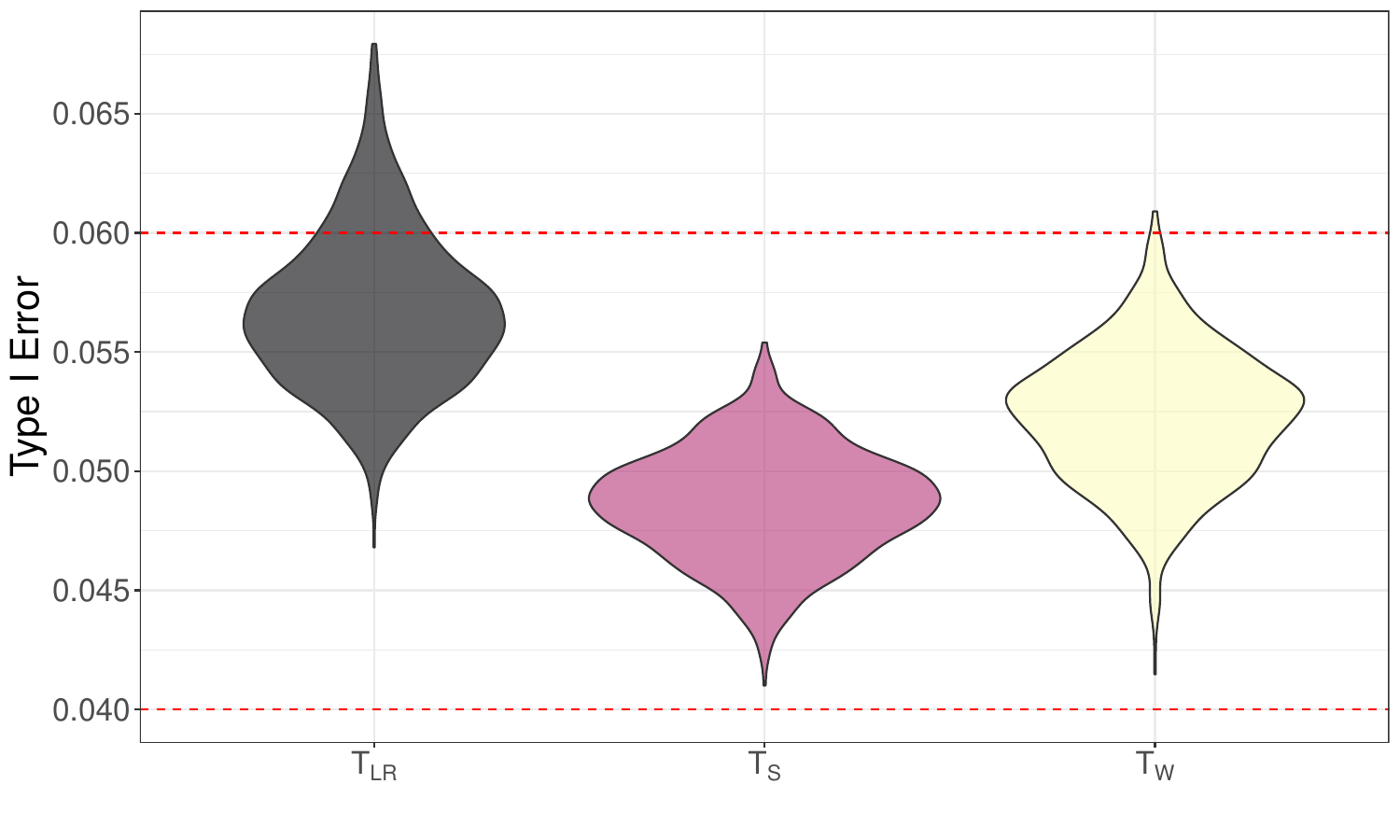}
        \caption{Violin Plots of Empirical Type I Errors for $g=6$ and $m_i=30$}
        \label{fig:violin_g6_m_30}
\end{figure}

Next, we assess the power performance of the proposed methods under numerous alternative hypotheses defined in \hyperref[tab:powerconf]{Table \ref*{tab:powerconf}}. The empirical powers when $g=3$ and 6 can be found in \hyperref[tab:power3]{Table \ref*{tab:power3}} and \hyperref[tab:power6]{Table \ref*{tab:power6}}, respectively.

\begin{singlespace}

\begin{table}[H]
\caption{Parameter Configurations for Empirical Powers}
\label{tab:powerconf}
\centering
\begin{tabular}{c||c|c}
\toprule

Case & Group number & $\pi_i$   \\ \hline
1&3 &$\pi_1=0.4$, $\pi_2=0.4$, $\pi_3=0.5$\\
2&3 &$\pi_1=0.4$, $\pi_2=0.4$, $\pi_3=0.53$\\
3&3 &$\pi_1=0.5$, $\pi_2=0.5$, $\pi_3=0.67$ \\
4&3 &$\pi_1=0.6$, $\pi_2=0.6$, $\pi_3=0.8$\\
A & 6& $\pi_1=0.4$, $\pi_2=0.4$, $\pi_3=0.45$, $\pi_4=0.45$, $\pi_5=0.5$, $\pi_6=0.5$\\
B & 6&$\pi_1=0.4$, $\pi_2=0.4$, $\pi_3=0.45$, $\pi_4=0.45$, $\pi_5=0.53$, $\pi_6=0.53$\\
C & 6&$\pi_1=0.5$, $\pi_2=0.5$, $\pi_3=0.6$, $\pi_4=0.6$, $\pi_5=0.67$, $\pi_6=0.67$\\
D & 6&$\pi_1=0.6$, $\pi_2=0.6$, $\pi_3=0.7$, $\pi_4=0.7$, $\pi_5=0.8$, $\pi_6=0.8$\\

\bottomrule
\end{tabular}
\end{table}
\end{singlespace}

\begin{singlespace}
\begin{center}
\addtolength{\tabcolsep}{-0.25em}
\begin{longtable}[H]{c c c | c c c | c c c | c c c }
\caption{The Empirical Powers (\%) when $g=3$ under the Nominal Level $\alpha=5\%$ } \label{tab:power3} \\

\hline
\multirow{2}{*}{$\theta$} & \multirow{2}{*}{Case} & Maximum of  & \multicolumn{3}{c|}{$m_i=30$} & \multicolumn{3}{c|}{$m_{i}=55$} & \multicolumn{3}{c}{$m_{i}=100$}  \\ \cline{4-12}
 &  & $|\pi_i-\pi_j|$, ($i \neq j$) & $T_{LR}$ & $T_{S}$ & $T_W$ & $T_{LR}$ & $T_{S}$ & $T_W$ & $T_{LR}$ & $T_{S}$ & $T_W$   \\  \hline

\endfirsthead

\multicolumn{12}{l}%
{{ \tablename\ \thetable{}: (Continued)}} \\
\hline
\multirow{2}{*}{$\theta$} & \multirow{2}{*}{Case} & Maximum of  & \multicolumn{3}{c|}{$m_i=30$} & \multicolumn{3}{c|}{$m_{i}=55$} & \multicolumn{3}{c}{$m_{i}=100$}  \\ \cline{4-12}
 &  & $|\pi_i-\pi_j|$, ($i \neq j$) & $T_{LR}$ & $T_{S}$ & $T_W$ & $T_{LR}$ & $T_{S}$ & $T_W$ & $T_{LR}$ & $T_{S}$ & $T_W$   \\  \hline
\endhead

\hline \multicolumn{12}{c}{{Continued on next page}} \\ \hline
\endfoot

\hline \hline
\endlastfoot

0 & 1 & 0.10 & 18.030 & 17.040 & 17.136 & 30.410 & 29.770 & 29.948 & 53.310 & 53.110 & 53.125 \\ 
        ~ & 2 & 0.13 & 28.120 & 27.370 & 27.526 & 49.650 & 49.140 & 49.307 & 77.290 & 77.110 & 77.211 \\ 
        ~ & 3 & 0.17 & 46.390 & 44.780 & 45.108 & 75.330 & 74.790 & 74.931 & 95.230 & 95.090 & 95.076 \\ 
        ~ & 4 & 0.20 & 69.180 & 67.250 & 67.533 & 92.960 & 92.450 & 92.473 & 99.770 & 99.740 & 99.737 \\ 
2 & 1 & 0.10 & 15.710 & 15.090 & 15.330 & 24.680 & 24.270 & 24.610 & 41.340 & 41.340 & 41.370 \\ 
        ~ & 2 & 0.13 & 22.800 & 22.150 & 22.700 & 37.970 & 37.770 & 37.800 & 62.560 & 62.730 & 62.590 \\ 
        ~ & 3 & 0.17 & 36.570 & 35.720 & 35.850 & 59.840 & 59.330 & 59.270 & 86.520 & 86.430 & 86.260 \\ 
        ~ & 4 & 0.20 & 53.755 & 52.025 & 51.875 & 80.830 & 80.010 & 79.900 & 97.510 & 97.410 & 97.340 \\ 
8 & 1 & 0.10 & 14.181 & 13.391 & 13.561 & 20.740 & 20.590 & 20.680 & 34.960 & 35.000 & 34.900 \\ 
        ~ & 2 & 0.13 & 20.082 & 19.592 & 19.702 & 32.460 & 32.390 & 32.450 & 54.170 & 54.060 & 54.100 \\ 
        ~ & 3 & 0.17 & 29.947 & 28.676 & 28.766 & 51.550 & 51.130 & 51.130 & 78.330 & 78.120 & 78.140 \\ 
        ~ & 4 & 0.20 & 45.598 & 43.827 & 43.707 & 71.960 & 70.640 & 70.650 & 93.980 & 93.660 & 93.660 \\

\end{longtable}
\end{center}
\end{singlespace}

\begin{singlespace}
\begin{center}
\addtolength{\tabcolsep}{-0.25em}
\begin{longtable}[H]{c c c | c c c | c c c | c c c }
\caption{The Empirical Powers (\%) when $g=6$ under the Nominal Level $\alpha=5\%$ } \label{tab:power6} \\

\hline
\multirow{2}{*}{$\theta$} & \multirow{2}{*}{Case} & Maximum of  & \multicolumn{3}{c|}{$m_i=30$} & \multicolumn{3}{c|}{$m_{i}=55$} & \multicolumn{3}{c}{$m_{i}=100$}  \\ \cline{4-12}
 &  & $|\pi_i-\pi_j|$, ($i \neq j$) & $T_{LR}$ & $T_{S}$ & $T_W$ & $T_{LR}$ & $T_{S}$ & $T_W$ & $T_{LR}$ & $T_{S}$ & $T_W$   \\  \hline

\endfirsthead

\multicolumn{12}{l}%
{{ \tablename\ \thetable{}: (Continued)}} \\
\hline
\multirow{2}{*}{$\theta$} & \multirow{2}{*}{Case} & Maximum of  & \multicolumn{3}{c|}{$m_i=30$} & \multicolumn{3}{c|}{$m_{i}=55$} & \multicolumn{3}{c}{$m_{i}=100$}  \\ \cline{4-12}
 &  & $|\pi_i-\pi_j|$, ($i \neq j$) & $T_{LR}$ & $T_{S}$ & $T_W$ & $T_{LR}$ & $T_{S}$ & $T_W$ & $T_{LR}$ & $T_{S}$ & $T_W$   \\  \hline
\endhead

\hline \multicolumn{12}{c}{{Continued on next page}} \\ \hline
\endfoot

\hline \hline
\endlastfoot

0 & A & 0.10 & 17.600 & 16.370 & 16.480 & 31.350 & 30.720 & 30.812 & 56.290 & 55.740 & 55.743 \\ 
        ~ & B & 0.13 & 29.206 & 27.576 & 27.932 & 52.730 & 52.230 & 52.492 & 82.830 & 82.660 & 82.745 \\ 
        ~ & C & 0.17 & 51.100 & 49.200 & 49.590 & 80.180 & 79.930 & 80.096 & 98.290 & 98.260 & 98.238 \\ 
        ~ & D & 0.20 & 74.870 & 73.220 & 73.433 & 96.310 & 96.050 & 96.164 & 99.980 & 99.980 & 99.980 \\ 
2 & A & 0.10 & 15.320 & 14.020 & 14.710 & 23.960 & 22.880 & 23.700 & 43.040 & 42.240 & 42.710 \\ 
        ~ & B & 0.13 & 23.360 & 21.800 & 22.420 & 40.120 & 39.200 & 39.620 & 68.450 & 68.040 & 68.180 \\ 
        ~ & C & 0.17 & 37.820 & 36.060 & 36.640 & 64.490 & 63.620 & 64.230 & 91.120 & 90.910 & 91.120 \\ 
        ~ & D & 0.20 & 57.506 & 55.586 & 55.916 & 86.680 & 86.180 & 86.290 & 99.010 & 98.980 & 98.990 \\ 
8 & A & 0.10 & 13.580 & 12.610 & 12.840 & 20.240 & 19.560 & 19.760 & 35.270 & 34.800 & 34.950 \\ 
        ~ & B & 0.13 & 19.430 & 18.300 & 18.560 & 34.170 & 33.440 & 33.540 & 58.870 & 58.490 & 58.550 \\ 
        ~ & C & 0.17 & 31.930 & 30.290 & 30.420 & 55.010 & 54.250 & 54.290 & 83.620 & 83.420 & 83.460 \\ 
        ~ & D & 0.20 & 48.277 & 46.013 & 46.023 & 77.260 & 76.340 & 76.320 & 97.180 & 97.070 & 97.070 \\

\end{longtable}
\end{center}
\end{singlespace}

\section{Data Analysis} \label{sec:data}

Two real-world examples are utilized to demonstrate the application of our proposed methods. The first is a cross-sectional study conducted in the Varamin district in Iran to assess the prevalence and causes of blindness and visual impairment (VI) \citep{rajavi2011rapid}. A total of 3,000  persons from 60 clusters were included using a multistage cluster systematic random sampling. Among them, 2819 persons were able to provide visual acuity measurements. The distribution of the available persons by age group and blindness status is exhibited in \hyperref[tab:eg1]{Table \ref*{tab:eg1}}. The maximum likelihood estimates using the proposed method of parameters can be found in \hyperref[tab:mle1]{Table \ref*{tab:mle1}}. The null hypothesis of homogeneous $\pi_i$ is rejected at $\alpha=5\%$ since $T_{LR}=136.589$ (p-value<0.0001), $T_S=178.749$ (p-value<0.0001), and $T_W=174.248$ (p-value<0.0001). The copula parameter estimate gives Kendall's tau around 0.6 indicating positive dependence in visual acuity measurements.

\begin{singlespace}

\begin{table}[H]
\caption{Example 1: Distribution of the Available Persons by Age Group and Blindness}
\label{tab:eg1}
\centering
\begin{tabular}{l c c c c c c c c}
\toprule

 \textbf{Blindness} & \textbf{50–54 yrs} & \textbf{55–59 yrs} & \textbf{60–64 yrs} & \textbf{65–69 yrs} & \textbf{70–74 yrs} & \textbf{75–79 yrs} & \textbf{$\geq$ 80 yrs}  & \textbf{Total}\\
\midrule
 None & 873& 541& 469& 257 &242 &127& 104& 2613\\
 Unilateral &23 &17 &18 &16 &32 &30 &29 &165 \\
 Bilateral & 2 &8 &4 &5 &3 &9 &10 &41 \\
\hline
 \textbf{Total} & 898 & 566 &491 &278 &277 &166 &143 &2819 \\

\bottomrule
\end{tabular}
\end{table}
\end{singlespace}

\begin{singlespace}

\begin{table}[H]
\caption{Maximum Likelihood Estimates of Parameters under $H_0$ and $H_a$ (Example 1)}
\label{tab:mle1}
\centering
\begin{tabular}{l c c c c c c c c }
\toprule

 \textbf{Hypothesis} &  \textbf{MLE} & \textbf{50–54 yrs} & \textbf{55–59 yrs} & \textbf{60–64 yrs} & \textbf{65–69 yrs} & \textbf{70–74 yrs} & \textbf{75–79 yrs} & \textbf{$\geq$ 80 yrs}  \\
\midrule
 & $\hat{\pi}_i$ & 0.015 &0.030& 0.027& 0.048& 0.067& 0.139&0.163\\
 $H_a$ & $\hat{\theta}$ &\multicolumn{7}{c}{4.581 }\\
  & $\hat{\rho}_i$ & 0.065 &0.120 &0.109& 0.180& 0.236& 0.395& 0.434 \\ \hline
 & $\hat{\pi}_{H_0}$ & \multicolumn{7}{c}{0.044 }\\
$H_0$ & $\hat{\theta}_{H0}$ & \multicolumn{7}{c}{9.740 }\\
 & $\hat{\rho}_{H0}$ & \multicolumn{7}{c}{0.301 }\\

\bottomrule
\end{tabular}
\end{table}
\end{singlespace}

Another example is from an observational study investigating the treatment effect of Orthokeratology (Ortho-k) on myopia at the First Affiliated Hospital of Xiamen University in 2023 \citep{liang2024homogeneity}. Ortho-k is a myopia correction method and is non-surgical. Subjects are required to wear specialized contact lenses overnight and the improvement of vision is temporary. There are two different lens designs. One is called corneal refractive therapy (CRT) and other is called vision shaping treatment (VST). Male subject distribution by design is presented in \hyperref[tab:eg2]{Table \ref*{tab:eg2}}. The maximum likelihood estimates of parameters can be found in \hyperref[tab:mle2]{Table \ref*{tab:mle2}}. We fail to reject the null hypothesis that $\pi_{VST}=\pi_{CRT}$ at the significant level of 5\% as $T_{LR}=0.034$ (p-value=0.8546), $T_S=0.034$ (p-value=0.8543), and $T_W=0.034$ (p-value=0.8539). The copula parameter estimate gives strong positive dependence in vision outcomes.

\begin{singlespace}

\begin{table}[H]
\caption{Example 2: Distribution of Male Subjects by Lens Design}
\label{tab:eg2}
\centering
\begin{tabular}{l c c c}
\toprule
 \textbf{No. of eyes with vision improvement} & \textbf{VST} & \textbf{CRT}  & \textbf{Total}\\
\midrule
 0 & 11& 6& 17 \\
 1 &4 &2 &6  \\
 2 & 3 &2 &5  \\
\hline
\textbf{Total} & 18 & 10 &28  \\

\bottomrule
\end{tabular}
\end{table}
\end{singlespace}

\begin{singlespace}

\begin{table}[H]
\caption{Maximum Likelihood Estimates of Parameters under $H_0$ and $H_a$ (Example 2)}
\label{tab:mle2}
\centering
\begin{tabular}{l c c c  }
\toprule

 \textbf{Hypothesis} &  \textbf{MLE} & \textbf{VST} & \textbf{CRT}  \\
\midrule
 & $\hat{\pi}_i$ & 0.276 & 0.303 \\
$H_a$ & $\hat{\theta}$ &\multicolumn{2}{c}{3.051 }\\
 & $\hat{\rho}_i$ & 0.466 & 0.491 \\ \hline
 & $\hat{\pi}_{H_0}$ & \multicolumn{2}{c}{0.286 }\\
 $H_0$ & $\hat{\theta}_{H0}$ & \multicolumn{2}{c}{3.050 }\\
 & $\hat{\rho}_{H0}$ & \multicolumn{2}{c}{0.475 }\\

\bottomrule
\end{tabular}
\end{table}
\end{singlespace}

\section{Discussion} \label{sec:discussion}
In this article, we proposed a new framework for testing the difference of event rates between groups with correlated paired data based on the Clayton copula. We investigated three hypothesis testing procedures and made likelihood inference. The optimization procedure is implemented using the \textsf{R} language. Simulation results indicated that all proposed testing procedures maintain satisfactory Type-I error control and exhibit reasonable power. Among them, the score test provides the best trade-off between  error control and power, regardless of the number of groups, sample size, or parameter configurations. On the other hand, the LR test has inflated Type-I errors when the sample size is small, which is acceptable because real studies rarely involve extremely small samples. As the sample size increases, the three test procedures began to perform more similarly.

Compared to previous studies based on classical methods, our work yields similar simulation results. Therefore, we conclude that using copulas to model dependencies in correlated paired data under this setting is feasible. The success of this framework lays a foundation for our future research. In upcoming studies, selecting an appropriate copula and exploring properties of binary margin copulas will become increasingly important. Furthermore, this approach offers a more general perspective on similar models.

In addition, for specific copula models, we can extend our inference to confidence intervals. For instance, \citet{tang2011asymptotic} investigated asymptotic confidence intervals for the difference in disease rates between two groups, i.e., $\pi_2 - \pi_1$. \citet{yang2021simultaneous} explored many-to-one simultaneous confidence intervals for comparing multiple treatments with a reference group. \citet{pei2012confidence} studied confidence intervals of proportion differences in a two-arm randomized clinical trial. Moreover, stratified designs can also be examined. For example, \citet{qiu2019tests, qiu2019construction} investigated asymptotic tests and confidence intervals for disease rates using stratified designs. \citet{shen2018testing, shen2019common} studied tests for the homogeneity of differences between two disease rates across strata and provided interval estimations of a common risk difference. Beyond bilateral-only studies, we can further generalize to bilateral-mixed-with-unilateral designs. Under this combined data framework, for example, \citet{ma2021testing} investigated asymptotic test methods for homogeneity of disease rates among multiple groups based on the ``constant R'' model.

\section*{Acknowledgment}
We would like to thank Dr. Guanjie Lyu for his review and suggestions.

\section*{Conflict of Interest}

All authors declare that they have no conflicts of interest.

\selectlanguage{english}

\bibliographystyle{apalike} 
\bibliography{ref.bib}

\appendix

\begin{appendices}
\section{Fisher Information Matrix}
The diagonal elements of the Fisher information matrix can be expressed as:\\

\[I_{ii}=E\Bigl(-\frac{\partial \ell^2 }{\partial \pi_i^2}  \Bigr)=\frac{A_i}{B_i},\]
where
\begin{flalign*}
A_i=&4m_i +{\left(8m_i +2m_i {{\left(1-\pi_i \right)}}^{2\theta } \right)}{{\left(2-{{\left(1-\pi_i \right)}}^{\theta } \right)}}^{2/\theta } +{\left(8m_i {{\left(1-\pi_i \right)}}^{\theta } -2m_i {\left({{\left(1-\pi_i \right)}}^{2\theta +1} +6\right)}\right)}{{\left(2-{{\left(1-\pi_i \right)}}^{\theta } \right)}}^{1/\theta }\\
&-8m_i {{\left(2-{{\left(1-\pi_i \right)}}^{\theta } \right)}}^{2/\theta } {{\left(1-\pi_i \right)}}^{\theta }  \\
B_i=&{\left({{\left(2-{{\left(1-\pi_i \right)}}^{\theta } \right)}}^{1/\theta } -1\right)}{{\left({{\left(1-\pi_i \right)}}^{\theta } -2\right)}}^2 {\left(\pi_i -1\right)}{\left(\pi_i -2\pi_i {{\left(2-{{\left(1-\pi_i \right)}}^{\theta } \right)}}^{1/\theta } +{{\left(2-{{\left(1-\pi_i \right)}}^{\theta } \right)}}^{1/\theta } -1\right)},\;\\ &(i=1,2,...,g)
\end{flalign*}
and 
\[I_{(g+1)(g+1)}=E\Bigl(-\frac{\partial \ell^2 }{\partial \theta^2}  \Bigr)=\frac{C_i}{D_i},\]
where
\begin{flalign*}
    C_i=&-m_i {{\left(2\log \left(-\frac{{{\left(1-\pi_i \right)}}^{\theta } -2}{{{\left(1-\pi_i \right)}}^{\theta } }\right)+2\theta \log \left(1-\pi_i \right)-\log \left(-\frac{{{\left(1-\pi_i \right)}}^{\theta } -2}{{{\left(1-\pi_i \right)}}^{\theta } }\right){{\left(1-\pi_i \right)}}^{\theta } \right)}}^2\\
    &\times {\left(3\pi_i {{\left(-\frac{{{\left(1-\pi_i \right)}}^{\theta } -2}{{{\left(1-\pi_i \right)}}^{\theta } }\right)}}^{1/\theta }  -{{\left(-\frac{{{\left(1-\pi_i \right)}}^{\theta } -2}{{{\left(1-\pi_i \right)}}^{\theta } }\right)}}^{1/\theta }  -2{\pi_i }^2 {{\left(-\frac{{{\left(1-\pi_i \right)}}^{\theta } -2}{{{\left(1-\pi_i \right)}}^{\theta } }\right)}}^{1/\theta }  +1\right)}\\
    D_i=&\theta^4 {{\left({{\left(1-\pi_i \right)}}^{\theta } -2\right)}}^2 \Biggl[3\pi_i {{\left(-\frac{{{\left(1-\pi_i \right)}}^{\theta } -2}{{{\left(1-\pi_i \right)}}^{\theta } }\right)}}^{1/\theta } -2{{\left(-\frac{{{\left(1-\pi_i \right)}}^{\theta } -2}{{{\left(1-\pi_i \right)}}^{\theta } }\right)}}^{1/\theta } -3\pi_i {{\left(-\frac{{{\left(1-\pi_i \right)}}^{\theta } -2}{{{\left(1-\pi_i \right)}}^{\theta } }\right)}}^{2/\theta } \\
    &+{{\left(-\frac{{{\left(1-\pi_i \right)}}^{\theta } -2}{{{\left(1-\pi_i \right)}}^{\theta } }\right)}}^{2/\theta }  +2{\pi_i }^2 {{\left(-\frac{{{\left(1-\pi_i \right)}}^{\theta } -2}{{{\left(1-\pi_i \right)}}^{\theta } }\right)}}^{2/\theta }  +1 \Biggr]. \\
\end{flalign*}
Elements at the bottom and right margins are:
\[I_{i(g+1)}=I_{(g+1)i}=E\Bigl(-\frac{\partial \ell^2 }{\partial \pi_i \partial \theta}  \Bigr)=\frac{E_i}{F_i},\]
where
\begin{flalign*}
E_i=&-2m_i {\left(2\log \left(-\frac{{{\left(1-\pi_i \right)}}^{\theta } -2}{{{\left(1-\pi_i \right)}}^{\theta } } \right)+2\theta \log \left(1-\pi_i \right)-\log \left(-\frac{{{\left(1-\pi_i \right)}}^{\theta } -2}{{{\left(1-\pi_i \right)}}^{\theta } } \right){{\left(1-\pi_i \right)}}^{\theta } \right)}\\
&\times \Biggl[ \pi_i {(-\frac{{{\left(1-\pi_i \right)}}^{\theta } -2}{{{\left(1-\pi_i \right)}}^{\theta } }) }^{1/\theta } -{(-\frac{{{\left(1-\pi_i \right)}}^{\theta } -2}{{{\left(1-\pi_i \right)}}^{\theta } }) }^{1/\theta } \\
&+\pi_i {(-\frac{{{\left(1-\pi_i \right)}}^{\theta } -2}{{{\left(1-\pi_i \right)}}^{\theta } }) }^{1/\theta } {{\left(1-\pi_i \right)}}^{\theta } -{\pi_i }^2 {(-\frac{{{\left(1-\pi_i \right)}}^{\theta } -2}{{{\left(1-\pi_i \right)}}^{\theta } }) }^{1/\theta } {{\left(1-\pi_i \right)}}^{\theta } +1 \Biggr]\\
F_i=&\theta^2 {{\left({{\left(1-\pi_i \right)}}^{\theta } -2\right)}}^2 {\left(\pi_i -1\right)} \Biggl[3\pi_i {{\left(-\frac{{{\left(1-\pi_i \right)}}^{\theta } -2}{{{\left(1-\pi_i \right)}}^{\theta } }\right)}}^{1/\theta } -2{{\left(-\frac{{{\left(1-\pi_i \right)}}^{\theta } -2}{{{\left(1-\pi_i \right)}}^{\theta } }\right)}}^{1/\theta } -3\pi_i {{\left(-\frac{{{\left(1-\pi_i \right)}}^{\theta } -2}{{{\left(1-\pi_i \right)}}^{\theta } }\right)}}^{2/\theta } \\
&+{{\left(-\frac{{{\left(1-\pi_i \right)}}^{\theta } -2}{{{\left(1-\pi_i \right)}}^{\theta } }\right)}}^{2/\theta }  +2{\pi_i }^2 {{\left(-\frac{{{\left(1-\pi_i \right)}}^{\theta } -2}{{{\left(1-\pi_i \right)}}^{\theta } }\right)}}^{2/\theta }  +1\Biggr], (i=1,2,...,g).\\
\end{flalign*}

\end{appendices}

\end{document}